# Hydra: AN ADAPTIVE–MESH IMPLEMENTATION OF P³M–SPH

H.M.P. Couchman

Department of Astronomy
University of Western Ontario
London, Ontario
N6A 3K7
Canada

P.A. Thomas and F.R. Pearce
Astronomy Centre, MAPS
University of Sussex
Falmer
Brighton
BN1 9QH
UK

## ABSTRACT

We present an implementation of Smoothed Particle Hydrodynamics (SPH) in an adaptive-mesh P³M algorithm. The code evolves a mixture of purely gravitational particles and gas particles. SPH gas forces are calculated in the standard way from near neighbours. Gravitational forces are calculated using the mesh refinement scheme described by Couchman (1991). The AP³M method used in the code gives rise to highly accurate forces. The maximum pairwise force error is set by an input parameter. For a *maximum* pairwise force error of 7.7%, the RMS error in a distribution of particles is $\approx 0.3\%$. The refined–mesh approach significantly increases the efficiency with which the neighbour particles required for the SPH forces are located. The code, "Hydra," retains the principal desirable properties of previous P³M–SPH implementations; speed under light clustering, naturally periodic boundary conditions and easy control of the accuracy of the pairwise interparticle forces. Under heavy clustering the cycle time of the new code is only 2–3 times slower than for a uniform particle distribution, overcoming the principal disadvantage of previous implementations — a dramatic loss of efficiency as clustering develops. A 1000 step simulation with 65 536 particles (half dark, half gas) runs in one day on a Sun Sparc10 workstation.

The choice of time integration scheme is investigated in detail. We find that a simple single-step Predictor–Corrector type integrator, which is equivalent to Leapfrog for velocity-independent forces, is the most efficient.

A method for generating an initial distribution of particles by allowing a a uniform temperature gas of SPH particles to relax within a periodic box is presented. The average SPH density that results varies by $\approx \pm 1.3$ percent. This is the fluctuation amplitude on roughly the Nyquist frequency, for smaller wavenumbers the fluctuations have lower amplitude. We present a modified form of the Layzer–Irvine equation which includes the thermal contribution of the gas together with radiative cooling.

The SPH and time integration schemes were tested and compared by running a series of tests of sound waves and shocks. These test were also used to derive timestep constraints sufficient to ensure both energy and entropy conservation.

We have compared the results of simulations of spherical infall and collapse with varying numbers of particles. We show that many thousands of particles are necessary in a halo to correctly model the collapse.

As a further test the cluster simulation of Thomas and Couchman (1992) has been re-run with the new code which includes a number of improvements in the SPH implementation. We find close agreement except in the core properties of the cluster which are strongly affected by entropy scatter in the older simulation. This demonstrates the crucial importance of conserving entropy in SPH simulations.

# 1. INTRODUCTION

Gasdynamical simulations are an indispensable tool for understanding the formation and evolution of cosmological structure. In this paper we present an N-body hydrodynamics code, Hydra, which uses a combination of Adaptive $P^3M$ for the gravity and Smoothed Particle Hydrodynamics for the gas forces. The requirements of the code are such that experiments with $\sim 5 \times 10^5$ particles (half dark, half gas) can be run for 1000 timesteps in a week on an ordinary 10Mflop workstation.

Smoothed Particle Hydrodynamics (SPH) has become a widely used tool in astrophysics. The fully Lagrangian formulation of SPH is particularly well suited to cosmic structure formation studies because it can adapt to the varying geometries and large density contrasts encountered in cosmological simulations. Although the accuracy of SPH does not compare well with other hydrodynamic methods with similar computational requirements, in particular with the Piecewise Parabolic Method (PPM; see, e.g., Collela & Woodward 1984), it is easier to implement and fits more naturally in particle-based gravity codes. Further, the wide dynamic range encountered in cosmological investigations would require refined grids of the PPM method in regions of high density. Although the gravity code upon which the present scheme is built uses refined grids it remains a considerable challenge to successfully match a grid-based method such as PPM to a cosmological particle-based gravity code. It is also true—although perhaps too seldom admitted—that SPH codes when used in the cosmological context are often pushed to the limit of the available resolution, at which point any hydrodynamic scheme may become little more than a convenient method for converting kinetic energy into heat; SPH offers a reliable way of achieving this.

SPH is straightforward to implement in gravity codes which already incorporate techniques allowing efficient searching over neighbouring particles. The two most popular gravity codes which have been used so far for astrophysical experiments with SPH are the $P^3M$ algorithm (Evrard 1990) and the Tree code (Hernquist & Katz 1989). The choice of which algorithm is most suitable for a particular application is largely dependent upon the characteristics of the gravity calculation in the code. The attractive features of the $P^3M$ algorithm are fast cycle time under light clustering, accurate interparticle forces with readily quantifiable errors, modest memory requirements and, for cosmological simulations, automatically periodic boundary conditions. The disadvantage of the $P^3M$ algorithm is its extremely slow cycle time under conditions of heavy particle clustering. For large numbers of particles the cycle time can become orders of magnitude slower than the time under light clustering. The Tree code is most suitable for the simulation of isolated objects and has the advantage that it does not suffer the same loss of efficiency as particle clustering develops. It is worth noting that a *fixed* smoothing length SPH algorithm implemented in the standard way as a direct accumulation of forces over near neighbours will suffer a degradation in performance under heavy clustering comparable to the $P^3M$ algorithm irrespective of which gravity solver is used.

The Adaptive $P^3M$ algorithm (Couchman 1991) retains all of the desirable features of $P^3M$ but avoids the severe time penalty associated with heavily clustered simulations. The poor performance of standard $P^3M$ arises when the number of particles in a grid cell becomes very large, resulting in the number of pairwise particle calculations over near neighbours becoming prohibitively large. The improvement of the $AP^3M$ algorithm is achieved by subdividing the grid in high-density regions into one or more finer grids (which may themselves be further subdivided) until there are just a few particles per grid cell. This technique offers an ideal framework for incorporating an SPH implementation in which the smoothing length is adjusted so that each particle interacts with a fixed number of neighbours.

One of our major goals is to maximise the efficiency of SPH–gravity schemes both in terms of minimizing the cycle time for the calculation of forces and in terms of memory requirement so that the maximum number of particles can be simulated within given computer hardware resources. In the cosmological context we are interested in the formation of clusters and ultimately galaxies.

A reasonably accurate representation of the shock heating of gas falling onto a spherical cluster requires more than a thousand particles. The resolution needed for a realistic cluster simulation will be several times this; the particle requirements for galaxy formation (in the cosmological context) will be many times higher still.

We begin in the following section by highlighting the details of the implementation of SPH in the current code which are special to the AP³M code as compared with the implementation of SPH in standard P³M. Section 3 describes tests of various algorithmic aspects of the code both for gravity and for SPH. We highlight and demonstrate the accuracy of the force calculation. Together with the technical details in Section 3 we present a new method for generating a quiet, but non-gridlike, initial particle distribution. These initial conditions are used in the tests of the code presented in later sections. In Section 4 we investigate in detail a range of time integration schemes with a view to optimising the efficiency of the code. These schemes are used in Sections 5 and 6 in tests of the SPH and in spherical infall tests respectively. In Section 7 we compare the results of a simulation of cluster collapse using both the new code and an older, non-adaptive version, and in Section 8 we summarise our conclusions and discuss the strengths of the new code and its areas of application.

## 2. IMPLEMENTATION

In this section we describe the implementation of the Hydra algorithm. Gravitational forces on the particles are calculated using the AP³M algorithm outlined by Couchman (1991). The essential details of this technique are described in Section 2.1. Gas forces are calculated using the standard SPH technique (e.g., Gingold & Monaghan 1977; Hernquist & Katz 1989) with some minor modifications as noted below. The fundamental point to remember is that both the gravitational and the gas forces calculated by Hydra are fully equivalent to those calculated by the straightforward implementation of SPH in a P³M code. The significance of this new technique is that it does not suffer the rapid slowdown under heavy clustering of the standard algorithm.

The implementation of SPH in the AP³M code falls into two parts; finding neighbouring particles and calculating the SPH forces. SPH forces are calculated by smoothing over the approximately 32 nearest neighbour particles. Thus in a high density region the smoothing kernel will extend over a smaller radius than in a region of low particle density. In the present code the refined meshes used for the gravity calculation in the AP³M algorithm in regions of high particle density allow the search for SPH neighbours to be restricted to a volume which more closely matches the SPH kernel. This is the primary difference between the SPH implementation described by Thomas & Couchman (1992) and Hydra. The details of this matching are described in the following subsection.

### 2.1 Neighbour search

We begin with a brief overview of the standard P³M algorithm and of the principal features of the AP³M algorithm. The purpose of P³M is to calculate high resolution gravitational forces. To this end the gravitational forces interpolated from a regularly spaced grid using a particle-mesh solver are augmented by a short range component summed from nearby particles. The mesh force has a softening of order the grid Nyquist wavelength ∼2 grid units. Thus the short range component needs to be accumulated from all particles closer than approximately two grid units. In a region of heavy clustering the range of the short range sum may enclose a large number of neighbours. (The precise radial cutoff distance for the short-range sum is determined by the required accuracy in the pairwise force as discussed in Section 3.)

The AP³M algorithm avoids the costly direct sum over near neighbours of standard P³M by replacing the direct sum in regions of high particle density with a P³M-type calculation performed on a refined mesh. The effect of this is to split the gravitational force due to near neighbours into a new mesh part followed by a further direct neighbour sum over a reduced search length. In very high density regions a series of refined meshes may be used which accumulate the forces on particles within the refinements to successively higher resolution. The final direct sum is then performed over a very much smaller search radius enclosing close to the number of neighbours optimal for efficient operation of the $P^3M$ algorithm (see below).

The effect of limiting the gravity direct sum to small search radii in high density regions is broadly consistent with the requirements of variable-smoothing-length SPH. The smoothing length is chosen so that the number of neighbours is approximately constant, typically 32. The refinement-placing criteria used by the $AP^3M$ algorithm is such that the average number of particles within the gravitational search radius for the final direct sum is in the range 4–32. (See Couchman 1991 for a discussion of the optimal number of particles per search volume.) Typically, for any centrally condensed object, the particle density will be higher in the centre of a refinement than at the edge. Indeed, it is advantageous in the $AP^3M$ algorithm to have a region of low particle density around the outside of a refinement. Unfortunately this means that the search radius required for gravity and that required for the SPH kernel will not always match. The implementation of SPH in Hydra ensures that neighbouring particles are searched for at the particular stage in the refinement hierarchy when the search is most efficient.

The $AP^3M$ algorithm accumulates forces by performing the following steps: i) the mesh force is calculated; ii) regions of high particle density are identified. These regions, termed refinements, will be covered by a mesh of a higher resolution; iii) the short range sum is performed in all regions not covered by a refinement. These three steps are then repeated in exactly the same way on each refined region until the list of refinements is exhausted. At this stage the force on each particle will have been fully accumulated. An example of the distribution of refinements is shown in Figure 1. This shows the gas particles in the cluster simulation described in Section 7 in a slice of half the depth of the simulation cube together with smaller boxes outlining the position of each of the refinements.

Calculation of the SPH quantities is performed during the calculation of the short range gravity sums in much the same way as occurs in standard $P^3M$. The full SPH force is accumulated on a particle from each of its neighbours at the same time. This is in contrast with the gravitational force which is accumulated in a number of steps from each of a succession of refinements and then from the final direct sum. At the stage when a refinement is solved in the $AP^3M$ algorithm all external forces have been previously accumulated and the particles in a refinement are treated as an isolated group with vacuum boundary conditions. There are two situations in which this assumption would lead to errors in the SPH forces on gas particles: if the smoothing volume for a particle overlaps the edge of the refinement or if the smoothing radius exceeds the size of the short range gravity search. In both these cases neighbours could potentially be missed. We avoid this problem by checking each particle which has been marked for further refinement to see if it will fail either of these conditions. If so the SPH forces are calculated at the (coarser) refinement level presently being calculated and the particle is flagged as having been done. The number of gas particles which fail either of the tests depends upon the local distribution of gas particles around the particle under consideration and the grid size relative to the local mean interparticle separation. The cost of implementing SPH in the $AP^3M$ algorithm in this way is a small amount of book-keeping to keep track of which particles have had SPH forces applied. The gain, however, is enormous; not only is the gravitational direct sum reduced to near optimal levels but the SPH forces on a gas particle are calculated at a stage in the refinement hierarchy when the minimum number of neighbours need be searched to accumulate the correct forces. The extra array which flags the status of a gas particle may also be conveniently used to identify which gas particles have become collisionless 'star' particles if star formation is included in the calculation.

Once the correct neighbour particles have been identified the accumulation of SPH forces proceeds as described in Thomas & Couchman (1992). During the loop which accumulates the gravitational force on an SPH particle, the density, pressure gradient and velocity divergence are evaluated. A temporary list of neighbouring particles lying within two smoothing lengths of the central particle is also generated. This list is then used to loop through the required neighbours a second time to apply equal and opposite SPH forces to the particle and each of its neighbours.

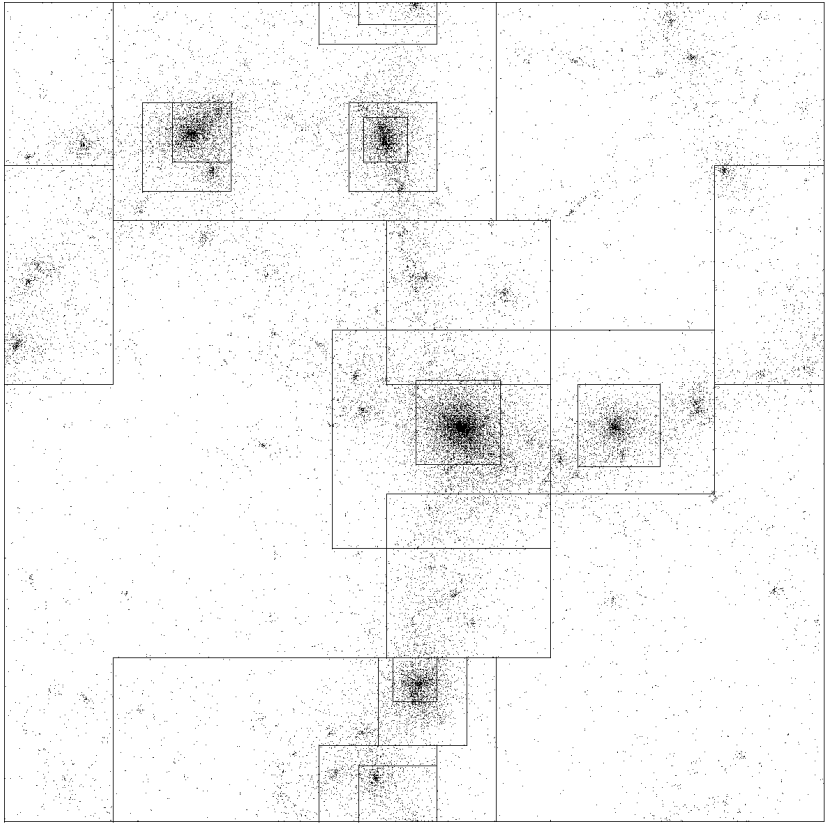

**Figure 1.** — The figure shows the distribution of refinements placed by the Hydra code for the final step of the cluster simulation described in Section 7. The gas particles in a projection of one half of the simulation cube are plotted. The apparent overlap of refinements is a result of the plotting projection. Small boxes within larger refinements are sub-refinements. The central cluster was refined four times in this simulation although for clarity only two levels are shown.

### 2.2 SPH forces

The SPH implementation is as described in Thomas and Couchman (1992) except that the rate of change of specific energy of particles is given by

$$\frac{d\epsilon_i}{dt} = -\frac{2}{3}\epsilon_i \nabla.\mathbf{v}_i$$

rather than

$$\frac{d\epsilon_i}{dt} = -\sum_j \frac{\epsilon_i}{m_i\epsilon_i + m_j\epsilon_j}(\mathbf{v}_i - \mathbf{v}_j).\mathbf{f}_{ij}.$$

Here $m_i$ is the mass, $\epsilon_i$ the specific energy and $\mathbf{v}_i$ the velocity of particle $i$, and $\mathbf{f}_{ij}$ is the force on particle $i$ due to particle $j$. The second of these formulae, whilst ensuring excellent energy conservation, induces a large scatter in entropy because of the pairwise estimation of the velocity divergence. This can lead to erroneous conclusions (see Section 7 for an example of this) and so we revert to the former, more usual equation.

Within a refinement which has a mesh spacing $\chi$ ($< 1$) times that of the original mesh we have the following scalings:

$$d\mathbf{r} \mapsto \chi\,d\mathbf{r}$$
$$h \mapsto \chi\,h$$
$$\mathbf{v} \mapsto \chi\,\mathbf{v}$$
$$n \mapsto \chi^{-3}n$$
$$\epsilon \mapsto \chi^{2}\epsilon$$

where $h$ is the SPH smoothing length and $n$ is the density (in units of particles per grid cell). It is then easy to verify that the pairwise gravity forces need to be scaled by a factor $\chi^{3}$ whereas the SPH forces are unaffected by the transformation. We have checked that the refined and unrefined versions of our code give identical SPH forces to within machine accuracy.

### 3. PERFORMANCE AND TECHNICAL ISSUES

In this section we discuss the performance of the Hydra code in two areas; force accuracy and cycle time. The pairwise force accuracy is directly measurable. The overall force on individual particles within a distribution of a large number of particles is, however, more difficult to measure. We estimate it by comparing forces returned by the code with and without refinements. In the same way the timestep efficiency of the code with refinements may be compared with one in which no refinements are placed. Switching off refinements in Hydra makes the code equivalent to the standard (SPH-)P$^3$M algorithm.

In addition to these tests it is convenient to present at this point a technique for generating initial conditions suitable for SPH and which is used in the SPH tests described in the following section.

### 3.1 *Pairwise gravitational force*

One of the principal strengths of P$^3$M is the ability to easily quantify and control the pairwise force. This feature is shared by the AP$^3$M algorithm. In order to understand the nature of the force error we start with a re-cap of the force calculation in the P$^3$M algorithm. Recall that the total force on a particle is composed of a long-range component calculated using a mesh based method, augmented by a short-range component summed directly from neighbouring particles. The sum of these components should reproduce the required force to some desired accuracy. The dominant error arises from the inability of the mesh to accurately represent fluctuations in the particle density on scales of order the mesh spacing. This error may be reduced by smoothing the particle distribution. The cost is that the short-range sum must then extend to larger radii. The larger search radius encloses more neighbour particles and, under heavy particle clustering, results in a dramatic rise in the computational effort required to calculate the short-range force.

The mesh smoothing function used in Hydra has compact support, thus the mean mesh force reproduces the required force down to a given radius, $r_m$. At smaller radii the mesh force falls below the required force as the mesh softening takes effect. The softening function used here is such that the mean mesh force reproduces the required force at significantly smaller radii than $r_m$ with only a small error. Since it is advantageous to reduce the range of the short-range sum, and since there is already an error present in the mesh force, we choose the cutoff radius for the short-range force to be such that the error between the *mean* mesh force and the required force at the cutoff radius is approximately equal to the error in the mesh force at that radius. The small discontinuity in the radial force which results is in practice swamped by the mesh force error as is evident from an inspection of Figure 2. At radii below the cutoff radius the short-range force is adjusted to exactly increment the mean mesh force to the value of the required force.

In the AP$^3$M algorithm (or Hydra) each level of refinement introduces a new mesh with a corresponding mesh error. Since the maximum pairwise mesh-force error occurs on scales near the mesh softening, each level of refinement effectively introduces an error at a different radial separation than its parent. The mesh softenings and hence short-range cutoffs for refinements are chosen so that the pairwise error for each mesh is approximately equal.

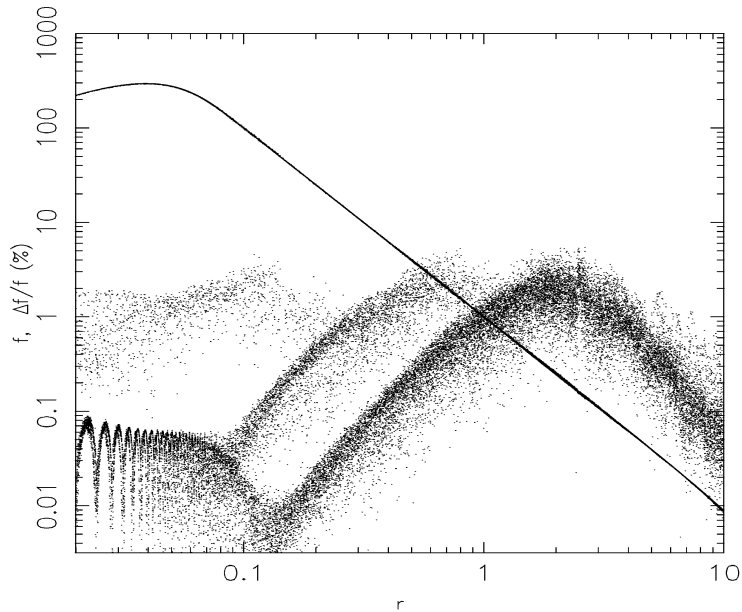

**Figure 2.** — The Pairwise gravitational force (continuous line) and fractional percentage error in the pairwise force (scattered dots) as a function of radial separation as measured in the Hydra code. The input accuracy parameter, perr, which sets the maximum error of this force was 7.7%. The error, $\Delta f$, is the modulus of the difference between the true force and the value measured from Hydra. As discussed in the text the errors are primarily the result of mesh-force errors. The three peaks in the error scatter plots represent, in order of decreasing radius, errors from the base mesh, a refinement of level one and from a daughter of that refinement at level two. Refinement meshes are chosen by the code so that the maximum error is the same at all levels refinement. It is evident that the RMS pairwise error is approximately 1–2% for this value of perr.

Figure 2 shows an example of the pairwise force and fractional error in the force for the base mesh and two levels of refinement. The errors from the meshes at three different levels are evident as the three peaks at different radii in the distribution of dots marking the fractional error in the force. It can be seen that three peaks are roughly the same height.

The forces are calculated by setting the mass of one particle to unity and measuring the accelerations of a number of massless surrounding particles. The massless test particles are distributed randomly in angle and with separation from the massive particle such that there are equal numbers in radial logarithmic bins. The experiment is then repeated with the massive particle moved to a new position with respect to the mesh and refinements. The comparison force has been calculated using the Ewald sum to correctly allow for the periodic boundary conditions at large separations (Ewald 1921 and see Efstathiou et al. 1985, Couchman 1991). The forces have been normalised to unity at unit base-mesh separation.

For practical purposes, a *maximum* fractional pairwise force error of order a few percent may be achieved. In the Hydra code the pairwise force error is selectable by an input parameter to the code. The requested maximum fractional error for the test illustrated in the figure was 7.7%. In principle any desired accuracy could be achieved with appropriate smoothing of the mesh force. In practice the code limits the requested accuracy to lie in the range 2–10%. A maximum force error similar to that shown in the figure is more than adequate for these simulations, as is evident from the figure the RMS error is very much less; of order 1%. Furthermore it is worth stressing again that the force error depicted is the pairwise error. The accuracy of forces in an ensemble of particles will be very much higher than this. As shown in the next subsection the RMS error in an ensemble of particles is of order 0.3% for a chosen maximum pairwise error of 7.7%.

### 3.2 Ensemble Forces

The force error on a particle in a distribution of particles will typically be very much less than the maximum pairwise force error shown in Figure 2 above. This was tested by comparing particle accelerations at a late stage of the cluster simulation with and without refinements. Under heavy clustering the particle accelerations are dominated in the unrefined case by the direct sum. Thus, although there is still an error from the base mesh, the comparison with accelerations calculated with refinements will give a good indication of the distribution of mesh-induced errors to be expected in an ensemble of particles. Acceleration differences are shown in Figure 3a for the dark matter and in Figure 3b for the SPH particles. A rough estimate of the accuracy of the ensemble forces may be made by comparing the RMS values of the acceleration and of the acceleration difference. For the dark matter the RMS acceleration in Figure 3 is 960 and the RMS difference in accelerations is 3.2, both values are in program units. The rms force error in this case is thus $\sim 0.3\%$.

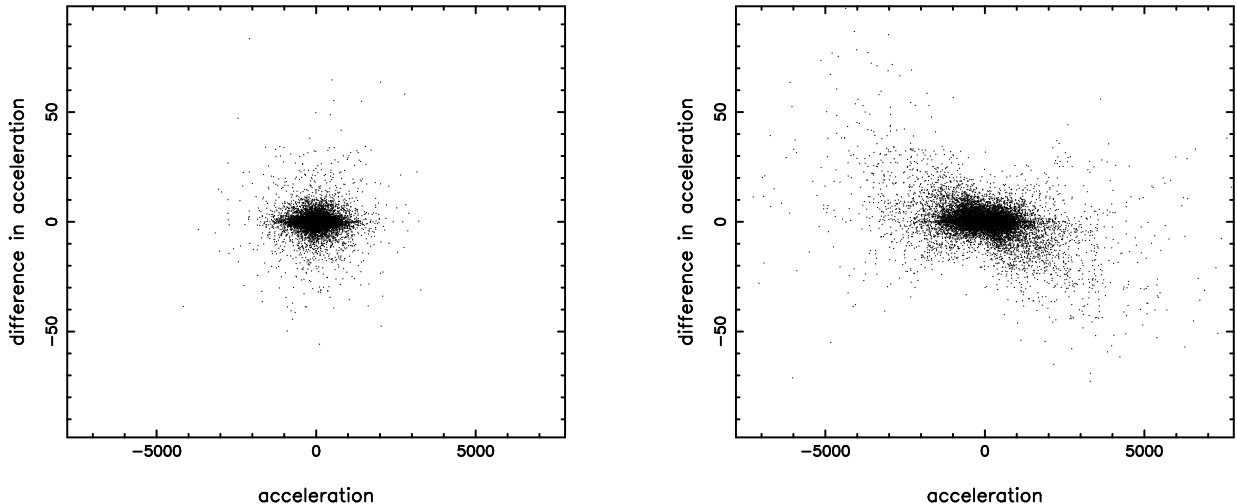

**Figure 3.** — Force differences for dark matter particles (a) and gas particles (b) when measured using Hydra with and without refinements. Note the very different scaling of the axes.

The improvement in efficiency in calculating SPH quantities in the Hydra code is simply the result of matching the smoothing length of an SPH particle to the appropriate level of refinement. Thus, the same neighbours are found for each SPH particle in both the refined and unrefined situations. Differences arise only because of roundoff error in scaling quantities to and from refinements. There is also the possibility, again from roundoff error, that a neighbour may be omitted from the neighbour list if its radial distance from the central particle is very close to the SPH search limit. Since the smoothing kernel is zero at this radius the associated error is negligible. It is worth contrasting this situation with the computation of the gravitational accelerations. In this case the neighbour particles which contribute to the final direct sum in the refined case will typically be a small subset of those that contribute in the unrefined case and most of the gravity will be accumulated using the mesh technique. This also means that the force on a particle near the edge of a refinement is typically composed of a refined-mesh component pointing into the refinement and a direct sum component pointing out of it.

An initial test was made of the accelerations on SPH particles by comparing the refined and unrefined cases with gravity turned off. As expected the accelerations were the same to within the limits of machine accuracy. The other SPH quantities — the density and temperature interpolated at each particle — are independent of the gravity calculation and were shown to be identical with and without gravity. Gravity was then switched on and the total accelerations compared for each set of particles; dark and SPH with the results as shown in Figure 3.

The distribution of errors for the SPH particles is broader with a more pronounced tail than for the dark matter. The particles with the largest fractional errors all have small overall accelerations which result from the near cancelation of accelerations accumulated at different levels of the gravitational calculation. It should be noted that these distributions will vary depending upon the degree of clustering in the particle distribution. They should be taken, therefore, only as being indicative of the overall level and distribution of errors.

### 3.3 Initial conditions

Several years ago we developed a method for producing a glass-like distribution of particles, of approximately uniform density and with small power on all scales. This may be used as an initial particle distribution before perturbation for a desired input fluctuation spectrum. A number of people have expressed interest in this technique and so we present the details here for the first time. We were motivated by the observation that a grid-like initial particle distribution is not a stable configuration in the presence of gas pressure forces. A particle distribution which is uniform on a cubic grid has the advantage that particle noise in absent, at least at early times, and the desired input spectrum may be reproduced almost exactly (Carlberg & Couchman 1989, Couchman 1994). Some concern has been expressed that a uniform initial distribution might lead to correlations in the final particle distribution which are unrelated to the input spectrum. This is probably a small effect, especially for initial spectra with small-scale power in which the grid structure is rapidly erased as the particle distribution evolves. The simplest alternative is to distribute particles randomly in the box. The disadvantage in this case is that shot noise will dominate the input spectrum on small scales.

As an alternative we created an initial distribution of particles by letting the box of SPH particles relax at constant entropy under the influence of gas pressure alone. The timestep was adjusted so as to produce motions of order the scatter in the interparticle separation per step. These motions were then damped by multiplying the velocities by a factor $< 1$ at the end of each timestep. A value of $3/4$ for the multiplicative factor was found to produce satisfactory results. This produces a pseudo-random distribution in which each particle has a roughly equal SPH density. In practice it is possible to achieve a distribution with a maximum variation in density of order $\pm 1.3\%$. The resulting power spectrum for this distribution is shown in Figure 4 for $32\,768\ (=32^3)$ particles in a cubic box. The corresponding level for Poisson noise is also indicated. It is evident that this technique reduces the fluctuation amplitudes in the initial particle distribution to no more than a few percent at all frequencies up to the effective Nyquist frequency of the particles — here a wavenumber of 16. (Note that the power spectrum in Figure 4 is plotted to twice the effective Nyquist frequency of the particles in order to illustrate the nature of the distribution on smaller scales.)

These initial conditions are used for all of the SPH tests described below. This is especially important for realistic shock tube experiments. Performing shock tubes in which particles are started on a grid, and hence collide head on in a shock, will results in shocks which are significantly better than those seen with a less structured initial particle distribution — such as would be encountered in a realistic gravitational collapse for example.

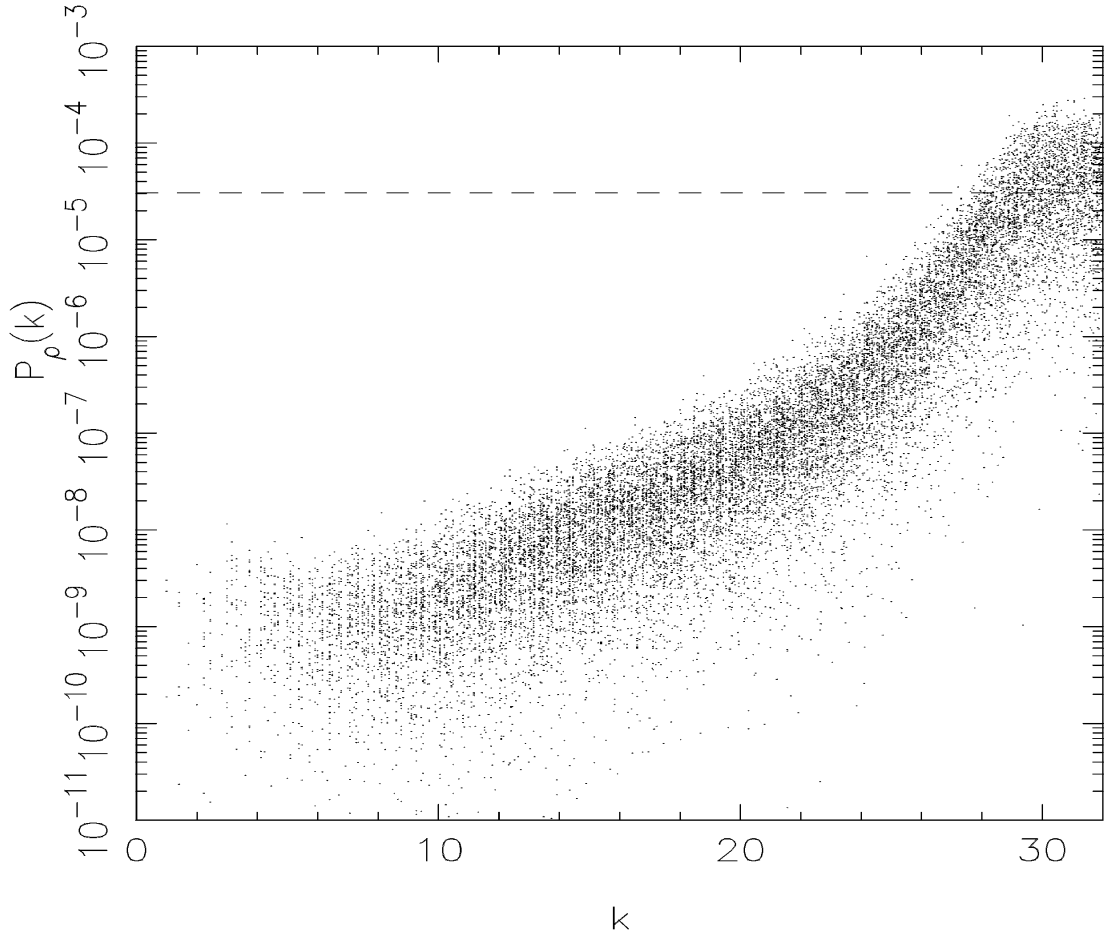

**Figure 4.** — The power spectrum of the relaxed initial particle distribution. The dashed line shows the Poisson noise level for 32 768 particles. The wavenumber is expressed as a multiple of the fundamental mode in the box plotted up to twice the Nyquist frequency of the particles. At the Nyquist frequency, $k = 16$, the mean power in the relaxed distribution is approximately three orders of magnitude below the level of Poisson noise.

### 3.4 *Timing*

The principle advantage of the adaptive mesh scheme described here over standard $P^3M$ is that the cycle time is proportional to the number of particles being simulated, independent of the degree of clustering of the particles. The cycle time under heavy clustering is a bounded multiple (of order 2.5) of the cycle time for the initial smooth particle distribution, whereas in the extreme limit the cycle time of standard $P^3M$ is bounded only by the time taken to perform the direct $N^2$ sum over all particles. (Timing issues for $AP^3M$ are discussed in detail by Couchman 1991.) Figure 5 shows the cycle time per step for the cluster simulation described in Section 7 for the standard and for the refined algorithms. At the end of the simulation the cycle time of Hydra has risen by a factor of 2.3 and is 3.7 times faster than the unrefined code.

In the purely gravitational code a region is refined when it is estimated that a saving in execution time will be achieved. The estimate is based upon the ratio of times taken to perform the mesh calculation and the direct sum. For the present code the mesh calculations are identical except that particles may have varying mass. The direct sum routine which now performs the gravity and SPH calculations is approximately 2 times slower than for gravity alone. The decision to place refinements has been adjusted to take account of the different relative times of the mesh and direct sum calculations but no further attempt has been made to allow for the different manner in which SPH neighbours and gravity neighbours are found. In principle it would be possible to tune the placing of refinements to allow the SPH density, and thus smoothing length, to dictate where refinements should be placed. Further improvement in efficiency may be possible by forcing more refinements at all levels so that the cutoff radius of the final direct sum is more closely matched to the SPH cutoff radius for a greater number of particles. The saving is unlikely to be substantial but has yet to be fully investigated. The routine which places refinements has been improved and optimized from the version discussed in Couchman (1991). The algorithm is now much better able to detect and adapt to clustering and mergers in hierarchical gravitational clustering. The success of this aspect of Hydra may be judged by the noting the relative smoothness of the refinement timestep curve in Figure 5.

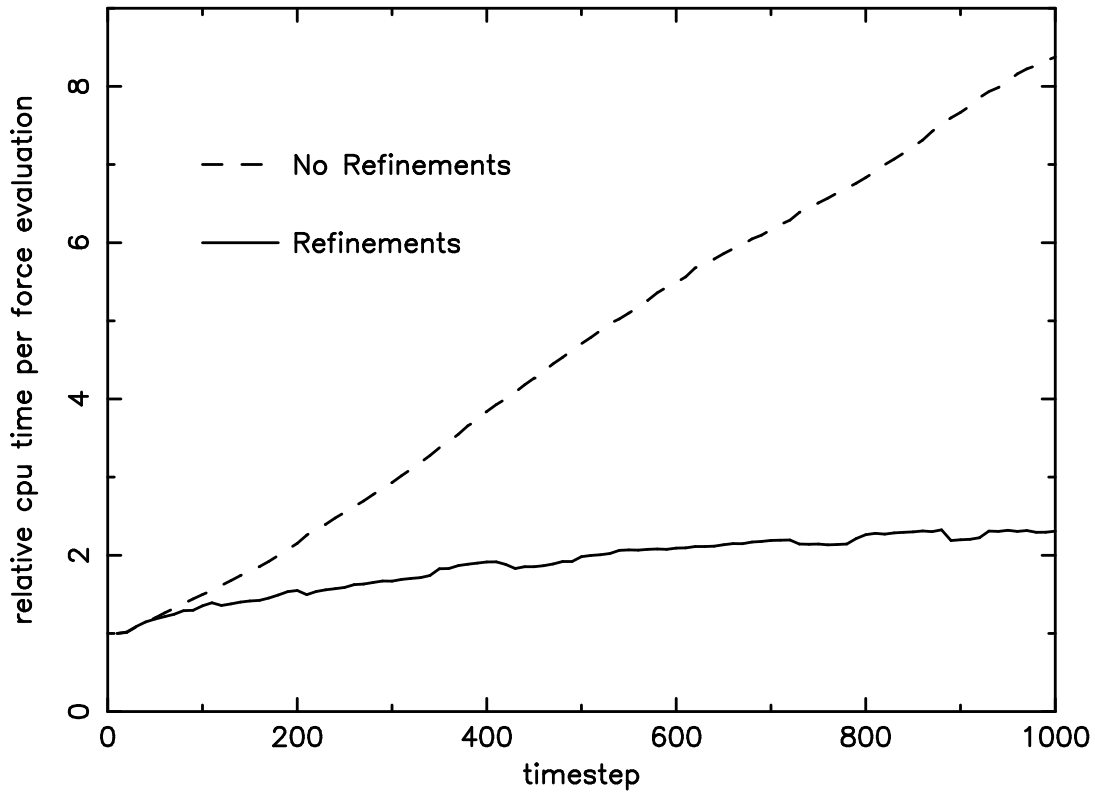

**Figure 5.** — Force evaluation time versus timestep for Hydra with and without refinements for the cluster simulation described in Section 7. Times are expressed relative to the time for the initial step.

In summary the fundamental performance difference between Hydra and the standard algorithm is that in standard $P^3M$ the time per step rises very rapidly as dense clusters develop, and this problem becomes worse as larger numbers of particles are used. In contrast for Hydra the increase in cycle time is bounded at $\sim 2.5$ times that for a uniform distribution and this ratio is independent of particle number.

## 4. TIME INTEGRATION SCHEMES

High resolution and hence large particle number is a fundamental goal of N-body cosmological simulations. Using very large numbers of particles results in a very high memory requirement and a force calculation which is very expensive in terms of CPU time. The overriding concerns in choosing a time integration scheme, therefore, are that it should minimize the amount of storage necessary for intermediate calculations and minimize the number of evaluations of the gravitational force necessary to achieve a specified accuracy over a given time interval.

Multistep (or equivalently multivalue) integration schemes require the storage of accelerations from previous timesteps. For each additional evaluation stored, the memory requirement is increased by $3N$ words for the accelerations alone. A further consideration is that the scheme should be well matched to the physical problem under investigation. When integrating the motion of SPH particles passing through a shock it may be inappropriate to use information from past timesteps. Further, it should be remembered that the stochastic nature of gravitational dynamics in many-body bound haloes imposes a limit on the accuracy that can be achieved for individual particle orbits. For these reasons we have concentrated on single step methods which can mostly be expressed as variations on Runge-Kutta (RK) schemes.

The integration schemes considered are described below. The schemes are shown assuming that the forces depend only upon particle positions and are independent of velocities. This is, of course, not correct for the SPH forces but will be sufficient to illustrate and compare the different schemes. A family of single step Runge–Kutta schemes for the integration of $\dot{\mathbf{v}} = \mathbf{f}$, $\dot{\mathbf{r}} = \mathbf{v}$ is given by:

$$\mathbf{r}_{n+1} = \mathbf{r}_n + \mathbf{v}_n \, dt + \left[ \left( 1 - \frac{1}{3\theta} \right) \mathbf{f}(\mathbf{r}_n) + \frac{1}{3\theta} \mathbf{f}(\mathbf{r}_\theta) \right] dt^2/2,$$

$$\mathbf{v}_{n+1} = \mathbf{v}_n + \left[ \left( 1 - \frac{1}{2\theta} \right) \mathbf{f}(\mathbf{r}_n) + \frac{1}{2\theta} \mathbf{f}(\mathbf{r}_\theta) \right] dt \tag{1}$$

where

$$\mathbf{r}_\theta = \mathbf{r}_n + \mathbf{v}_n \theta \, dt + \mathbf{f}(\mathbf{r}_n)(dt\theta)^2/2,$$

$\theta$ is a parameter and $dt$ is the timestep. We have considered $\theta = 1/2$, $2/3$ and $1$.

A second, related set of schemes is constructed by evaluating forces only once per step at predicted positions $\mathbf{r}'$. This force together with that saved from the last step are then used to correct the positions and velocities as below.

$$\mathbf{r}_{n+1} = \mathbf{r}_n + \mathbf{v}_n \, dt + \left( (1 - \alpha)\mathbf{f}(\mathbf{r}'_n) + \alpha\mathbf{f}(\mathbf{r}'_{n+1}) \right) dt^2/2,$$

$$\mathbf{v}_{n+1} = \mathbf{v}_n + \left( (1 - \beta)\mathbf{f}(\mathbf{r}'_n) + \beta\mathbf{f}(\mathbf{r}'_{n+1}) \right) dt \tag{2}$$

where

$$\mathbf{r}'_{n+1} = \mathbf{r}_n + \mathbf{v}_n \, dt + \mathbf{f}(\mathbf{r}'_n) \, dt^2/2,$$

This is like a single step Predictor–Corrector scheme and we label it PEC — **P**redict, **E**valuate, **C**orrect — in the usual manner. Using this terminology the above $\theta = 1$ Runge–Kutta scheme corresponds to EPEC (which is equivalent to the more familiar PECE Predictor–Corrector). The difference is that in the PEC scheme forces are only ever calculated at the predicted positions and will therefore contain information from the previous step. We attempt to determine below if the saving in evaluations outweighs any decrease in accuracy. We require $\beta = 1/2$ for the velocities to be accurate to second order. The choice of $\alpha$ is somewhat arbitrary. $\alpha = 0$ gives a scheme similar to the popular Leapfrog scheme (discussed below) except that the velocities are predicted forward to the same time as the positions before force evaluation. (In Leapfrog proper the velocities and positions are evaluated half a timestep apart.) The other case we consider is $\alpha = 1/3$ which nominally gives third-order accuracy for the positions (but this is swamped by the error in the velocities).

The 'Leapfrog' scheme is:

$$\mathbf{r}_n = \mathbf{r}_{n-1} + \mathbf{v}_{n-1/2}\, dt$$
$$\mathbf{v}_{n+1/2} = \mathbf{v}_{n-1/2} + \mathbf{f}(\mathbf{r}_n)\, dt \tag{3}$$

To allow for a variable timestep one simply replaces $dt$ by the mean value for adjacent timesteps in the latter of these equations, although this reduces the accuracy. Leapfrog integration has the advantage of minimizing storage: for forces which depend only upon position new values of both velocities and positions may overwrite the old values. If the forces depend upon velocities, as in SPH, then $\mathbf{v}_n$ must be estimated: using $\mathbf{v}_{n-1/2}$ decreases the accuracy of the scheme appreciably. The alternative is to predict $\mathbf{v}_n$ by incrementing $\mathbf{v}_{n-1/2}$ by half a step and then perform the Leapfrog. It is easily verified from equations (2) and (3) that this procedure is formally equivalent to the $\alpha = 0$ PEC scheme.

The truncation errors for each scheme are given in Table 1.

Table 1
Errors* and Stability

| Scheme | | $\epsilon_{local}$ | $\epsilon_{global}$ | $\gamma$ |
|---|---|---|---|---|
| | | $\mathbf{r}$ $\quad$ $\mathbf{v}$ | $\mathbf{r}, \mathbf{v}$ | |
| RK | $\theta = 1/2$ | 5 $\quad$ 3 | 2 | 2.25 |
| | $\theta = 2/3$ | 4 $\quad$ 4 | 3 | 2.14 |
| | $\theta = 1$ | 4 $\quad$ 3 | 2 | 2 |
| PEC | $\alpha = 0$ | 3 $\quad$ 3 | 2 | 2 |
| | $\alpha = 1/3$ | 4 $\quad$ 3 | 2 | 1.55 |
| Leapfrog | | 3 $\quad$ 3 | 2 | 2 |

*Figures recorded are $n$, where the error is $O(h^n)$

Each integration scheme was analysed in the standard manner (see, for example, Hockney & Eastwood, 1988, §4.4; Lapidus, 1976) for stability, or the growth of errors. This imposes a constraint on the timestep as a necessary condition for correct integration of stable solutions. If the desired solution is oscillatory with angular frequency $\Omega$ then we require $\Omega\, dt \leq \gamma$ where $\gamma$ is shown in the final column of Table 1. For growing solutions this constraint on the timestep is too restrictive: it is necessary only that the errors grow no faster than the desired solution, a condition which is satisfied by all the schemes considered here. This analytic investigation does not, of course, guarantee stability of the solutions for the particular problem under investigation but is a useful indicator of the relative stability of the schemes.

Expressing the equations of motion in comoving coordinates introduces an effective drag term due to the expansion of the background. This has the effect of increasing the stability of the integration scheme. This effect was not included in the stability analysis and is not relevant for the SPH tests considered below.

The relative performance of the integration schemes will be investigated in the next two sections.

## 5. SPH TESTS

In this section we discuss various hydrodynamic tests of the code. The code is run with SPH only (*i.e.*, no gravity) on standard problems such as sound waves and shock tubes and thus provides a stringent test of the relative merits of various time integration schemes and of the accuracy attainable by the code.

### 5.1 Relaxed Particle Distribution

The simplest and yet an important test of the SPH is that a uniform distribution of particles with only small random motions should remain uniform with no change in energy or entropy. To test this we started with a relaxed distribution of particles, as described in Section 3.3, where the temperature was chosen so that the sound speed, $c_s$, was equal to unity (in normalised code units). This gives a Courant timescale of $dt_C \equiv \min(h/c_s) \approx 1.0$. Integration schemes typically require the timestep to be less than some fraction of the Courant timescale for stability, for example Hernquist & Katz (1989) take $dt \lesssim 0.3 dt_C$ when using the Leapfrog integrator. We can derive a timestep constraint from the stability analysis of the previous section. The minimum wavelength for oscillations is two particle separations, or about $2h$, which gives an angular frequency of $\pi c_s/h = \pi/dt_C$. Hence we require $dt \leq (\gamma/\pi) dt_C \approx 0.6\text{–}0.7 dt_C$.

Table 2 shows the results of allowing the relaxed box to run on for 50 time units with a fixed timestep, as indicated.

Table 2
Relaxed distribution

| Scheme | $dt$ | $dt$/eval | $dE/E$ | $\sigma_S/S$ |
|---|---|---|---|---|
| RK $\theta = 1/2$ | 1.4 | 0.7 | $\infty$ | $\infty$ |
|  | 1.0 | 0.5 | $4.0 \times 10^{-4}$ | $3.1 \times 10^{-3}$ |
|  | 0.6 | 0.3 | $3.0 \times 10^{-4}$ | $1.8 \times 10^{-3}$ |
| $\theta = 2/3$ | 1.4 | 0.7 | $\infty$ | $\infty$ |
|  | 1.0 | 0.5 | $3.2 \times 10^{-4}$ | $2.7 \times 10^{-3}$ |
|  | 0.6 | 0.3 | $3.1 \times 10^{-4}$ | $1.5 \times 10^{-3}$ |
| $\theta = 1$ | 1.4 | 0.7 | $\infty$ | $\infty$ |
|  | 1.0 | 0.5 | $\infty$ | $\infty$ |
|  | 0.6 | 0.3 | $2.5 \times 10^{-4}$ | $1.4 \times 10^{-3}$ |
| PEC $\alpha = 0$ | 0.7 | 0.7 | $4.0 \times 10^{-4}$ | $1.4 \times 10^{-3}$ |
| (corrected | 0.5 | 0.5 | $3.5 \times 10^{-4}$ | $1.2 \times 10^{-3}$ |
| Leapfrog) | 0.3 | 0.3 | $3.3 \times 10^{-4}$ | $1.1 \times 10^{-3}$ |
| $\alpha = 1/3$ | 0.7 | 0.7 | $3.1 \times 10^{-4}$ | $1.2 \times 10^{-3}$ |
|  | 0.5 | 0.5 | $3.2 \times 10^{-4}$ | $1.1 \times 10^{-3}$ |
|  | 0.3 | 0.3 | $2.8 \times 10^{-4}$ | $1.2 \times 10^{-3}$ |

For the three RK runs with $dt = 1.4$ the divergence is exponential. The RK($\theta = 1$) run with $dt = 1.0$ remained stable for about 80 timesteps before diverging when the Courant timescale dropped below about 0.93. We conclude that the limiting ratio of $dt/dt_C$ for stability is just larger than unity for this scheme, slightly greater than the value estimated above. In every other case the energy increases linearly at a rate of $dE/E \approx 1\text{–}2 \times 10^{-6}$ per evaluation. We define an entropy-related quantity $S = T/n^{2/3}$ where $T$ is temperature and $n$ the density at the location of each particle. The mean entropy was approximately conserved in most cases but the fractional root-mean-square spread in entropy, $\sigma_S/S$, varied considerably.

The PEC schemes clearly come out best in this test. The energy conservation and spread in entropy are approximately constant, independent of the timestep. Both PEC algorithms are more stable than the RK ones for the same number of evaluations per time interval. In principal the RK with $\theta = 2/3$ could provide more accurate results if we made the timestep sufficiently small but in practice we are unlikely to want to do so.

### 5.2 Sound waves

The next test we tried was that of a sound wave in a cuboidal box of dimensions $6 \times 6 \times 168$, with periodic boundary conditions. Once again we started with relaxed initial conditions and particles were then displaced sinusoidally along the long-axis of the box by an amount $\delta z = (\pi s/2\omega)\sin kz$ where $\omega = kc_s$, $k = 6\pi/168$, and $s$ is a constant which measures the amplitude of the wave. This gives three full waves of wavelength 56 units along the long axis of the box. The theoretical period of oscillation is $T = 2\pi/kc_s = 56$ units. We ran the simulations for 100 time units corresponding to approximately 2 complete periods. Figure 6 shows an example of the measured oscillation in kinetic energy (two oscillations per period). These are well-fit by expressions of the form

$$K = K_0 e^{-t/t_d}\left(1 + e^{-t/t_w}\cos 4\pi t/T\right),\qquad(4)$$

where $t_d$ and $t_w$ are the respective timescales on which the motion is dissipated (conversion of kinetic energy into heat) and loses its coherence (either by particle interpenetration or because sin waves do not oscillate in perfect synchrony). The values of the parameters are similar in all cases; for example, the PEC($\alpha = \frac{1}{3}$) scheme with $dt = 0.5$ gives $T = 55.4$, $t_d = 272$ and $t_w = 160$. The measured period is close to (though slightly less than) the predicted one.

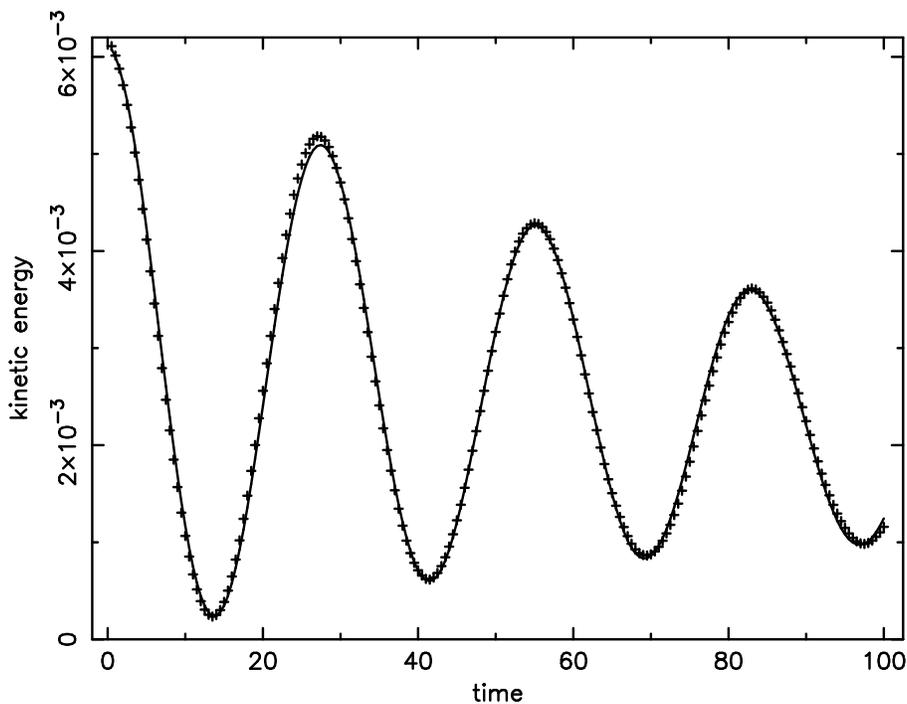

**Figure 6.** — Kinetic energy as a function of time for a sound wave with strength $s = 0.1$. The line is a fit of the function in Equation 4.

The changes in energy and entropy after 100 time units are shown in Table 3 for the various timestepping schemes. Note that we show here the maximum entropy spread which corresponds to slightly more than a 3 sigma deviation.

<div align="center">

Table 3
Sound Waves

</div>

| Scheme | $dt$/eval | Strength, $s = 0.1$ | | Strength, $s = 0.2$ | |
|---|---|---|---|---|---|
| | | $dE/E$ | $\Delta S/S$ | $dE/E$ | $\Delta S/S$ |
| RK $\theta = 1/2$ | 0.5 | 0.008 | 0.054 | | |
| | 0.3 | 0.001 | 0.029 | | |
| | 0.1 | $8 \times 10^{-5}$ | 0.023 | | |
| $\theta = 2/3$ | 0.5 | 0.004 | 0.046 | $\infty$ | $\infty$ |
| | 0.3 | $9 \times 10^{-4}$ | 0.027 | 0.0024 | 0.062 |
| | 0.1 | $6 \times 10^{-5}$ | 0.022 | $2 \times 10^{-4}$ | 0.058 |
| $\theta = 1$ | 0.5 | $\infty$ | $\infty$ | | |
| | 0.3 | 0.002 | 0.026 | | |
| | 0.1 | $9 \times 10^{-5}$ | 0.023 | | |
| PEC $\alpha = 0$ | 0.5 | 0.002 | 0.030 | 0.006 | 0.064 |
| (corrected | 0.3 | $5 \times 10^{-4}$ | 0.025 | 0.001 | 0.058 |
| Leapfrog) | 0.1 | $2 \times 10^{-4}$ | 0.021 | $6 \times 10^{-5}$ | 0.060 |
| $\alpha = 1/3$ | 0.5 | 0.001 | 0.025 | 0.003 | 0.058 |
| | 0.3 | $4 \times 10^{-4}$ | 0.027 | 0.001 | 0.059 |
| | 0.1 | $3 \times 10^{-5}$ | 0.021 | $8 \times 10^{-5}$ | 0.059 |

We conclude that the RK and PEC schemes have similar performances for a sufficiently small timestep. For larger timesteps the PEC schemes once again win out. The results for a stronger sound wave, $s = 0.2$, bear out this conclusion. In contrast to the relaxed test the PEC($\alpha = 1/3$) scheme performs slightly better than PEC($\alpha = 0$), but there is little to choose between them.

Once again it is the Courant condition which limits the maximum timestep for stability. $dt_\mathrm{C} = 0.89$ and $0.87$, respectively in each of the two set of sound waves. The results seem to imply a stability limit of $dt$/eval/$dt_\mathrm{C} \approx 0.6$ for the RK schemes.

### 5.3 Shock tubes

Finally, we report on the results of shock tube experiments. Once again these were carried out in a long cuboid of cross-section $6 \times 6$ grid units with a particle distribution relaxed independently on either side of the shock front. We tried two examples: a mild shock with density ratio two and entropy jump $S = 1.2$ to $S = 1.323$, and a strong shock with density ratio four and entropy jump $S = 0$ to $S = 0.198$. In each case the velocity difference across the shock was set equal to unity in code units.

In every case we get approximately the correct jump conditions with a shock width of about 6 particles in both density and velocity. The entropy profile is somewhat steeper and has a width of only 3–4 particles. An example of the measured SPH quantities for a strong shock are shown in Figure 7. The mean value of the entropy downstream of the shock and its root-mean-square deviation are shown in Table 4.

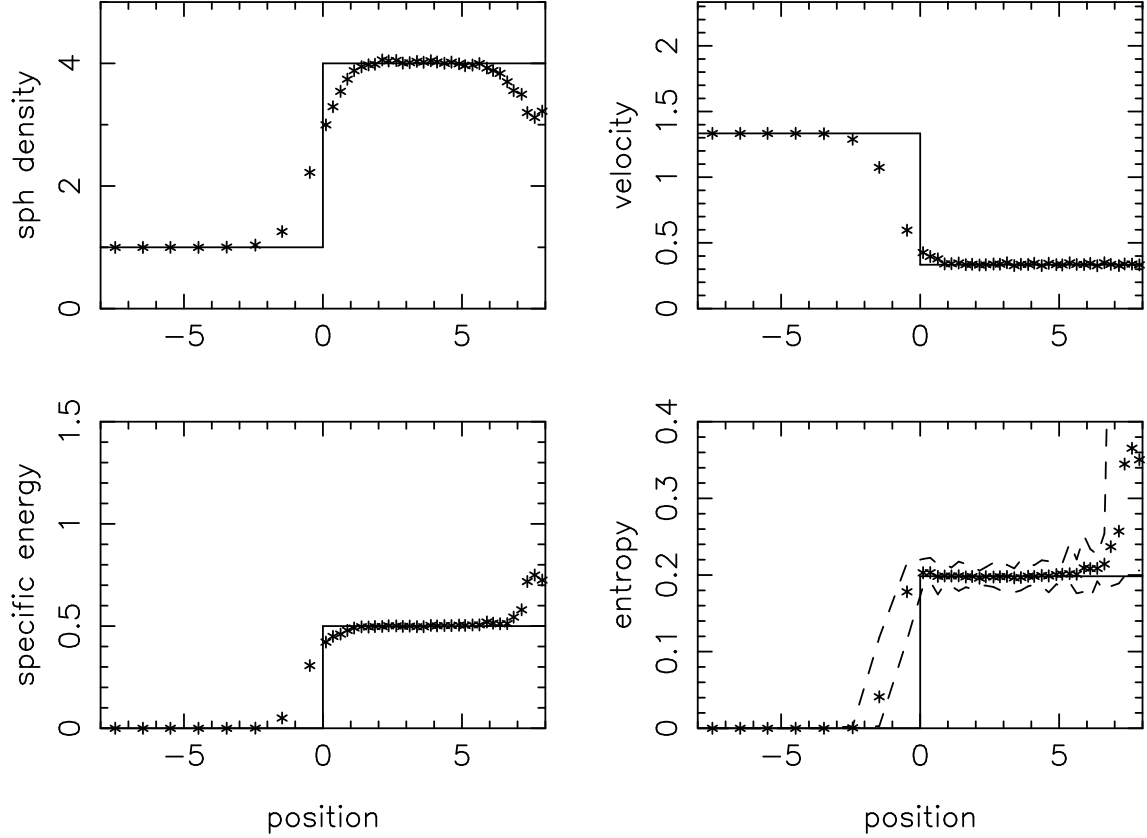

**Figure 7.**—Shock jump conditions for the strong shock described in the text. In each of the four panels the solid line indicates the theoretical jump across the shock. The dashed lines in the entropy panel indicate the maximum and minimum spread of the entropy. The fluctuations on the extreme right are from particles which started at the shock boundary.

Table 4
Shocks

| Scheme | $dt/\text{eval}$ | $\Delta\rho/\rho = 2$ | | $\Delta\rho/\rho = 4$ | |
|---|---|---|---|---|---|
| | | $\bar{S}$ | $\sigma_S$ | $\bar{S}$ | $\sigma_S$ |
| RK $\theta = 2/3$ | 0.4 | $\infty$ | $\infty$ | $\infty$ | $\infty$ |
| | 0.3 | $\infty$ | $\infty$ | 0.1973 | 0.0076 |
| | 0.2 | 1.320 | 0.010 | 0.1940 | 0.0083 |
| PEC $\alpha = 0$ | 0.4 | 1.325 | 0.009 | 0.1962 | 0.0067 |
| | 0.3 | 1.319 | 0.010 | 0.1938 | 0.0073 |
| | 0.2 | 1.316 | 0.010 | 0.1926 | 0.0085 |
| PEC $\alpha = 1/3$ | 0.4 | $\infty$ | $\infty$ | 0.1961 | 0.0066 |
| | 0.3 | 1.319 | 0.010 | 0.1939 | 0.0077 |
| | 0.2 | 1.316 | 0.010 | 0.1925 | 0.0087 |

It is interesting to note that longer timesteps give better entropy jump conditions and a smaller dispersion. This suggests that the artificial viscosity is not converting all the kinetic energy into heat. However the worst (non-divergent) case shown in the table has an entropy jump which is only 6 percent too low, which is sufficient for our purposes.

The required restrictions on the timestep are slightly more severe than for the previous tests. The measured values of $dt_C$ for the weak and strong shock simulations are about 0.50 and 0.78, respectively. Thus divergence is expected when $dt/\text{eval}/dt_C \approx 0.5$ (RK) or 0.7 (PEC $\alpha = 1/3$). It is not clear, however, that the Courant condition is the limiting factor in this situation. Two other relevant timescales are

$$dt_a \equiv \min_i (h_i/|a_i|)^{1/2}$$

$$dt_v \equiv \min_{i,j} (h_i/|\Delta v_{i,j}|),$$

where $a$ is the acceleration, $\Delta v$ the relative velocity, and the subscripts $i$ and $j \neq i$ run over all particles. $dt_a$ measures the timescale over which particles are being accelerated, and $dt_v$ the timescale for changes in geometry due to relative motion of particles. (Similar expressions for $dt_a$ and $dt_v$ can be written for the dark matter particles, simply by replacing $h$ with the maximum of the softening and the interparticle separation, whereas $dt_C$ obviously has no meaning for the dark matter.) We choose to keep $dt_C$, $dt_a$ and $dt_v$ separate rather than combining them together to form one grand timestep constraint.

In the strong shock we measure values of $dt_a \geq 0.8$ and $dt_v \geq 0.5$ which suggests that acceptable values of $dt/dt_a$ are $< 0.4$ (RK) and $< 0.5$ (PEC) and of $dt/dt_v$ are $< 0.6$ (RK) and $< 0.8$ (PEC). The weak shock results are less restrictive, but confirm that the PEC($\alpha = 0$) scheme has a slightly larger range of stability than PEC($\alpha = 1/3$).

### 5.4 Summary

In conclusion we find that good results are obtained from our sound wave and shock tube tests provided that the timestep is kept sufficiently small. For the same number of evaluations per timestep the PEC schemes seem to be more accurate and more stable than the RK schemes. We would expect the higher order schemes to do better as the timestep is reduced but this is unlikely to be acceptable for most applications. In the rest of the paper we concentrate on just two schemes: RK($\theta = 2/3$), the most accurate of the Runge-Kutta schemes which we tried, and PEC($\alpha = 0$) which is equivalent to the Leapfrog integrator for gravity-only simulations. We impose a timestep constraint $dt/\text{eval}/dt_C < 0.4$ and 0.5 for the two schemes, respectively. Other constraints will be discussed in more detail below.

## 6. SPHERICAL INFALL TESTS

### 6.1 Tophat collapses

To test the gravitational forces and energy integration scheme under conditions which resemble those found in simulations of structure formation, we study the collapse of overdense, uniform spheres embedded within a uniformly expanding background. The analytic solution to this problem is the well known top-hat solution. The density within the sphere remains uniform and its radius varies with time as a cycloid. We performed three simulations each with $N^3$ particles, where $N = 8$, 16 and 32, respectively.

The initial conditions for these simulations were constructed as follows. We start with a relaxed distribution of particles (see Section 3.3) corresponding to a uniform distribution. All particles within a sphere of radius 0.45 times the box-size were given initial displacements and velocities appropriate to a tophat with overdensity 0.1, while those outside the sphere were left unaltered. This radius encloses 38% of the particles.

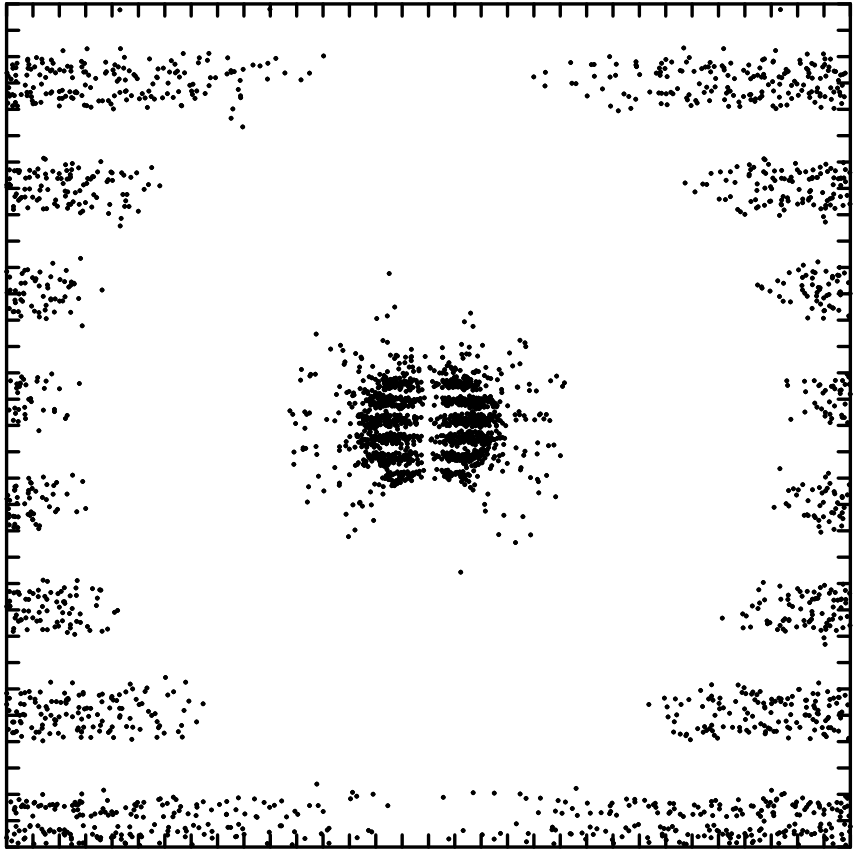

**Figure 8.** — The top hat collapse at a central overdensity of 180. A subset of the $32^3$ particles lying in horizontal sheets are shown projected by rotation into the plane of the page.

We look at two things: the uniformity of the collapse provides a check on the gravitational force, while the error in the shape of the cycloid tests the ability of the integration scheme to follow such highly unstable collapses. In each case the inner regions of the sphere collapse uniformly but the surface layer of particles gets left behind. This is evident in Figure 8 which shows a cross-section of the collapse of a $32^3$ box at the time when the overdensity is 180. The particle positions have been rotated about a vertical axis until they lie in the plane. We have omitted some of the particles in a series of horizontal sheets so that it is possible to see how these layers are preserved during the collapse.

Figure 9 shows the density profiles for each of the three boxes at the time when the overdensity is approximately 180. There is no surviving constant-density region in the smallest box. For the other two the density is constant (within statistical fluctuations) near the centre and then declines sharply at larger radii. In each case there is almost exactly one layer of particles in this surface region. It arises because of the irregularity of the potential due to the finite number of particles: any particles which lie at slightly larger radii than the average will feel a smaller acceleration than they should and so will 'peel away' from the collapsing sphere. This shows that in order to *accurately* model a collapsing dark matter halo very large numbers of particles are required. Of course it may be possible to obtain a close approximation to the correct density distribution in the final object with a smaller number of particles, and most realistic collapses will involve a high degree of substructure which would be more important than the shot-noise in the particle distribution. However this brings into question published results on the structure of galactic halos in which the objects in question contain only a few hundred particles.

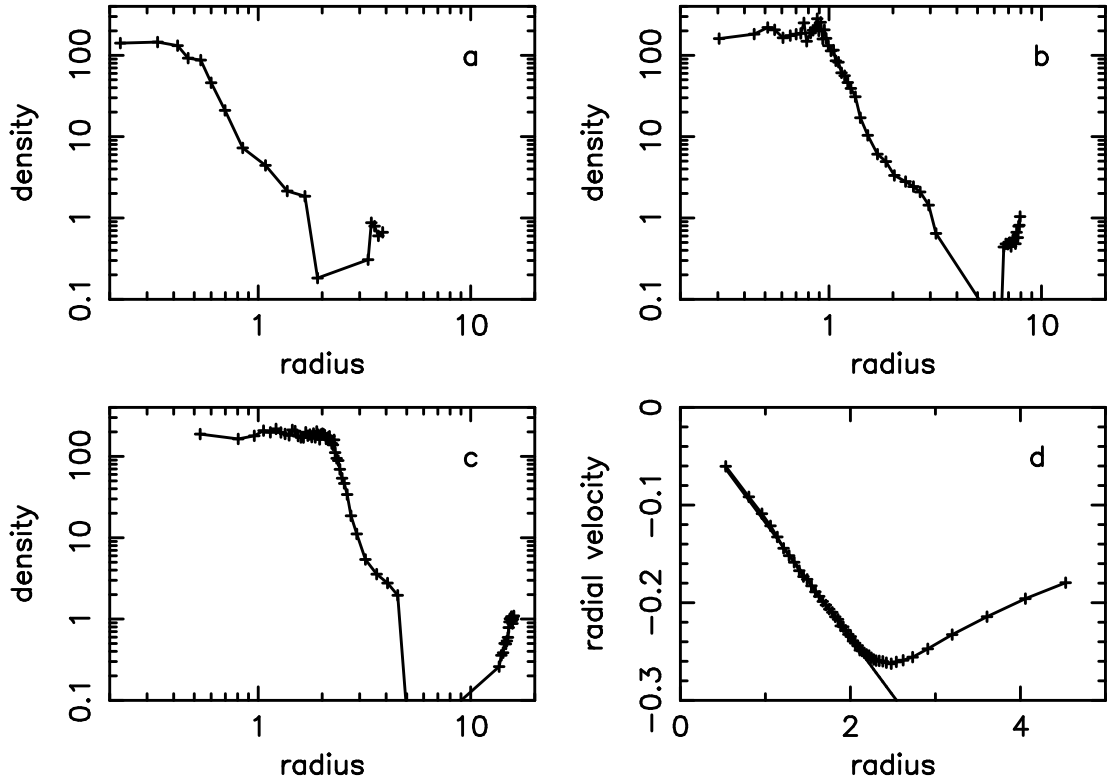

**Figure 9.** — Density profiles for top hat collapses with (a) 512, (b) 4096 and (c) 32 768 particles. Panel d) shows the radially averaged infall velocity for the 32 768 particle collapse.

Next we look at the rate of collapse. This can be measured either by the time at which the central overdensity reaches 180, or by the slope of the velocity-radius relation at this stage: both give similar results. We express the results in terms of the initial overdensity which would give the same collapse time or velocity/radius ratio if the theoretical tophat collapse had been perfectly followed:

Table 5
Tophat models

| N | $dt/dt_a$ | $\delta_0$ |
|---|---|---|
| 8 | 0.25 | — |
|   | 0.125 | — |
| 16 | 0.25 | 0.1013 |
|    | 0.125 | 0.1013 |
| 32 | 0.25 | 0.1013 |
|    | 0.125 | 0.1013 |
| 16 | 0.5 | 0.1020 |
|    | 0.25 | 0.1016 |

The first column in the Table gives the cube root of the number of particles, the second is the constraint imposed on the timestep. The final column gives the initial value of the overdensity which would give the same degree of collapse for the inner regions of the sphere at the time when the overdensity has reached 180 (recall that the initial overdensity was 0.1). The collapses were run with two values of the maximum pairwise force error (see Section 3); 7.7% and 2% but this made little difference to the results. There is no entry for $N = 8$ as the irregularities in the surface layer have penetrated right to the core by this stage in the collapse. The upper set of results in each pair is for the RK($\theta = 2/3$) timestepping routine, and the lower using PEC($\alpha = 0$) with the same total number of evaluations. The two give similar results except for large timesteps when PEC performs slightly better. We conclude that the rate of growth of overdensity is adequately followed for all choices of timestepping algorithm which give reasonable energy and entropy conservation.

### 6.2 Secondary infall

The top-hat density profile gives a singular collapse in which all the particles pass through the origin at the same time. We wish to test for energy conservation in a collapse which is more closely related to those which we might expect in our simulations. To do this we again chop a sphere out of an otherwise uniform box, but this time we arrange for an overdensity profile $\delta \propto r^{-0.5}$. Thus the inner regions of the sphere will collapse first followed by shells of matter at successively larger radii. Once again the velocity perturbations are determined by the growing mode of the perturbation.

We measure energy conservation in expanding co-ordinates by the use of a modified form of the Layzer–Irvine Equation (Peebles 1980). The energy equation in expanding co-ordinates is

$$\frac{d}{dt}(K + U + W) + \frac{\dot{a}}{a}\big(2(K + U) + W\big) = -L$$

where $K$, $U$ and $W$ are the kinetic, thermal and potential energies, respectively, $L$ is the radiation rate (luminosity) of the gas, and $a$ is the expansion factor of the Universe. This can be turned into a variety of integral forms the most obvious of which is

$$I \equiv K + U + W + \int \big(2(K + U) + W\big)\frac{da}{a} + \int L\,dt = 0.$$

We use $|I/W|$ as a measure of the accuracy of our integrations.

### 6.2.1 Dark matter collapses: gravity only

We first look at gravity-only (*i.e.* dissipationless) collapses. The depth of the collapse is measured by the potential energy which peaks when the outer shell of material crosses the centre of the sphere. The percentage error as determined by $|I/W|$ peaks at this time (except in the case of small numbers of particles) and declines thereafter.

Table 6 shows the results for a variety of simulation parameters. The first three columns give respectively the cube root of the number of particles, the maximum percentage error in the pairwise gravitational force calculation, and the timestepping control parameter, $dt/dt_a$. Next come the maximum depth of the potential energy, the value of the Layzer–Irvine integral, $|I|$, at this time, and finally the percentage error defined by the ratio of these two. The runs are divided into four groups (separated by horizontal rules in the table) with different parameters varying in each group as indicated. The pair of runs with $N = 16$, perr=7.7% and $dt/dt_a = [0.25, 0.125]$ is repeated in each group for clarity. Within each group simulations are shown in pairs; the upper run in each case is performed using the RK($\theta = 2/3$) timestepping routine, and the lower using PEC($\alpha = 0$). The number of evaluations per time interval is equal for each member of a pair; i.e., two force evaluations per step for RK and one for PEC.

| N | perr | $dt/dt_a$ | $|W|_{max}$ | $|I|_{max}$ | % error |
|---|------|-----------|-------------|-------------|---------|
| 16 | 7.7 | 1.0 | $2.05 \times 10^4$ | $9.4 \times 10^2$ | 4.6 |
| | | 0.5 | $2.02 \times 10^4$ | $2.2 \times 10^2$ | 1.1 |
| | | 0.5 | $2.23 \times 10^4$ | $7.1 \times 10^2$ | 3.2 |
| | | 0.25 | $2.14 \times 10^4$ | $1.9 \times 10^2$ | 0.9 |
| | | 0.25 | $2.10 \times 10^4$ | $3.0 \times 10^2$ | 1.4 |
| | | 0.125 | $2.14 \times 10^4$ | $1.7 \times 10^2$ | 0.8 |
| | | 0.125 | $2.07 \times 10^4$ | $1.5 \times 10^2$ | 0.7 |
| | | 0.0625 | $2.12 \times 10^4$ | $1.5 \times 10^2$ | 0.7 |
| 16 | 7.7 | 0.25 | $2.10 \times 10^4$ | $3.0 \times 10^2$ | 1.4 |
| | | 0.125 | $2.14 \times 10^4$ | $1.7 \times 10^2$ | 0.8 |
| | | 2.0 | $2.12 \times 10^4$ | $1.8 \times 10^2$ | 0.7 |
| | | | $2.11 \times 10^4$ | $9.7 \times 10^1$ | 0.3 |
| 8 | 7.7 | 0.25 | $7.89 \times 10^2$ | $1.9 \times 10^{1†}$ | 2.4 |
| | | 0.125 | $7.96 \times 10^2$ | $8.8^†$ | 1.1 |
| 16 | | | $2.10 \times 10^4$ | $3.0 \times 10^2$ | 1.4 |
| | | | $2.14 \times 10^4$ | $1.7 \times 10^2$ | 0.8 |
| 32 | | | $5.29 \times 10^5$ | $2.8 \times 10^3$ | 0.5 |
| | | | $5.44 \times 10^5$ | $1.6 \times 10^3$ | 0.3 |
| 16 | 7.7 | 0.25 | $2.10 \times 10^4$ | $3.0 \times 10^2$ | 1.4 |
| | | 0.125 | $2.14 \times 10^4$ | $1.7 \times 10^2$ | 0.8 |
| 16 gas | | | $2.30 \times 10^3$ | $1.2 \times 10^1$ | 0.5 |
| | | | $2.30 \times 10^3$ | $1.1 \times 10^1$ | 0.5 |
| 32 | | | $5.29 \times 10^5$ | $2.8 \times 10^3$ | 0.5 |
| | | | $5.44 \times 10^5$ | $1.6 \times 10^3$ | 0.3 |
| 32 gas | | | $1.97 \times 10^5$ | $2.0 \times 10^2$ | 0.1 |
| | | | $1.98 \times 10^5$ | $2.0 \times 10^2$ | 0.1 |

† $|I|$ continues to increase at later times in these models.

Energy conservation improves, as expected, with decreasing timesteps but this variation is much more marked for the RK scheme. For the runs with the PEC integrator in the first block of Table 6 ($N = 16$, perr$= 7.7\%$, $dt/dt_a \leq 0.25$) the percentage error is in the range 0.7–0.9 at the time of maximum collapse and then remains approximately constant. For $dt/dt_a = 0.5$ the percentage error is slightly higher and continues to increase after maximum collapse. (Some idea of the intrinsic errors associated with the accuracy of the potential calculation may be gained by noting that the measured potential energy jumps by approximately 0.01 percent on the introduction of the first refinement.) Overall energy conservation can be improved to 0.3 percent by lowering perr to 2.0 (in this case the integration takes 1.5 times as long). We estimate that this is the accuracy with which we calculate the conserved energy integral and so the true error in the integration may be even lower. For the RK scheme, 0.7 percent accuracy can only be achieved by taking a very small timestep, or once again setting perr$= 2.0$: accuracy is rapidly lost as the timestep is increased, and the percentage error continues to increase slowly after maximum collapse.

Energy conservation also varies with the number of particles. The percentage error decreases as more particles are used because small-scale fluctuations in the gravitational potential are relatively less important. The simulation with 512 ($N$=8) particles (of which about 200 are in the collapsing sphere) performs poorly. Energy is conserved to a few percent at the time of maximum collapse but the fractional error continues to grow at later times.

When we carried out this set of simulations we initially hoped that the final density profiles would reflect the self-similar nature of the collapse. For an overdensity profile $\delta \propto r^{-n}$ it is relatively easy to show that successive shells of infalling matter should end up with characteristic radii and densities which scale as $\rho \propto r^{-3n/(1+n)}$ ($n = 0$ is the tophat collapse, $n = 3$ is Bertschinger's 1985 secondary infall onto a point mass). Unfortunately this density profile is not apparent in our results (see Figure 10a). The reason for this is that each shell of matter virialises to form a complex density profile which must be convolved with the above power law in order to get the correct distribution. We simply do not have enough scale-lengths in our box for the power law to dominate (which will happen when each shell of matter occupies a small fraction of the total radius and so can be treated as a delta function in the convolution).

### 6.2.2 Gaseous collapses: gravity + SPH

Gaseous collapses differ from dark-matter ones in that the collapse is not as deep because the infalling gas is shocked at the outer edge of the virialised region and does not penetrate all the way down into the core. The potential energy increases monotonically, gradually levelling off after an initial steep rise, rather than peaking and then declining. For this reason the energy conservation is better. We show the results for the standard run, at the time when the corresponding dark matter has reached maximum collapse, at the foot of Table 6. The error in the energy integral is 0.5 percent for both integration schemes at this stage. It subsequently remains constant for the PEC scheme but continues to rise slowly for RK.

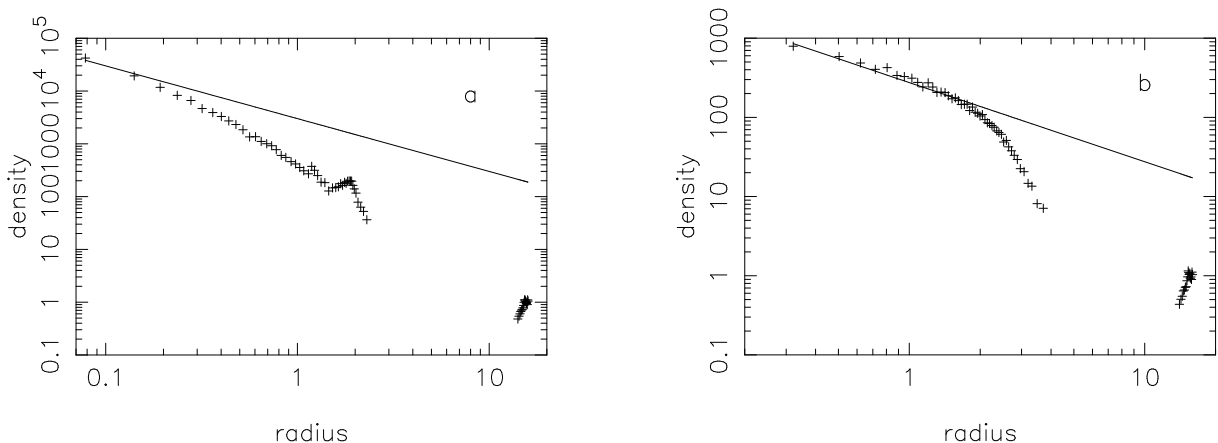

**Figure 10.** — Density profiles for the secondary infall tests; a) dark matter and b) gas. The solid line has a slope of -1 corresponding to the self similar infall solution for an initial density profile of $\rho \propto r^{-0.5}$.

The self-similar nature of the collapse is much better represented than for the collisionless case. Figure 10b shows the density profile at a time when the system has settled down into a quasi-steady state. In the inner part of the profile the density is inversely proportional to radius, as predicted. Only in the outer regions does the density decline due to the finite size of the simulated region. A velocity profile shows that the infall velocity is inversely proportional to radius and is just such as to balance the Hubble flow—in other words the distribution is correctly contracting in code units so as to maintain a constant physical size.

### 6.3 Summary

The code performs well in these collapse simulations. In particular it can conserve energy to a fraction of a percent in the secondary infall runs which most closely resemble the type of collapse we expect in the real Universe. The PEC integration scheme again comes out on top. In achieves a similar sort of accuracy to the RK scheme for small timesteps but unlike the RK it maintains this accuracy as the timestep is increased. The accuracy obtained depends upon both the length of timestep used as well as the value of the maximum pairwise force error, perr. To achieve a specified accuracy it may be more efficient to reduce perr, rather than to reduce the timestep. Not only is the integration intrinsically more accurate, but the integration time is increased by only a moderate amount.

As a result of all these studies we recommend the low-order, PEC($\alpha = 0$) integration scheme. This not only minimises storage but it is also the most accurate and the most stable for equivalent numbers of evaluations per timestep. For gravity-only runs this scheme is equivalent to the Leapfrog integrator which can be implemented without the need for any acceleration arrays. In order to obtain one percent accuracy with minimum entropy scatter during the collapse and virialisation of halos we adopt the following constraints: for the timestep $dt < \min(0.5\,dt_{\mathrm{C}},\, 0.25\,dt_{\mathrm{a}})$, and for the maximum pairwise gravitational force error, perr= 7.7. For higher accuracy perr should be reduced.

## 7. CLUSTER FORMATION

In order to test our code on conditions closely resembling those we wish to study, we have repeated an earlier simulation of the formation of a cluster performed using non-adaptive P$^3$M+SPH (Thomas & Couchman 1992, hereafter TC92). The simulation contained $32^3$ particles each of dark matter and gas, the former being nine times the mass of the latter.

We begin by outlining the relevant results from the previous experiment. The cluster formed by the merger of three large subgroups at a redshift of $z = 0.55$ and then grew by accretion of smaller subgroups and intergalactic material. At the end of the simulation there were approximately 2000 of each kind of particle within the virialised part of the cluster, twice this within the cluster as a whole. The density profiles of the gas and dark matter closely resembled one another and were well-fit by the formula

$$n(r) \propto \frac{1}{a^2 + r^2} - \frac{1}{b^2 + r^2} \qquad (5)$$

where $a = 50\,\mathrm{kpc}$ and $b = 1.25\,\mathrm{Mpc}$. The gravitational softening length in the simulation was 40 kpc and so there was no evidence of a core in either the gas or dark matter distributions. We estimated the 2-body relaxation time in the core of the cluster and found it to be less than a Hubble time. The core regions may therefore be affected by numerical relaxation and should not be considered as representative of real clusters.

The simulation using the Hydra code gives very similar results except in one important respect, which we will discuss below. The cluster forms at the same time and grows in the same way as before. The dark matter distribution at the end of the simulation is very similar to the P$^3$M+SPH one. The upper set of points in Figure 11 shows the density profile of dark matter binned in spherical shells about the position of the density maximum. The solid line shows the expected distribution using the above formula, which was derived from a fit to the gas density profile in the P$^3$M+SPH run. This figure should be compared to Figure 9 of TC92.

The equivalent gas profiles are shown in the lower set of points and by the lower solid curve. The Hydra code clearly gives a much more extended profile with a larger core radius. It is reasonably well-fit by the dotted curve which has the parameters $a = 120\,\mathrm{kpc}$ and $b = 2.5\,\mathrm{Mpc}$. The reason for the difference between the two runs is better illustrated by comparing the temperature and entropy profiles in Figure 12 with those of Figure 6 in TC92. The Hydra run has a much higher core entropy, and hence a much less precipitous temperature drop towards the centre of the cluster.

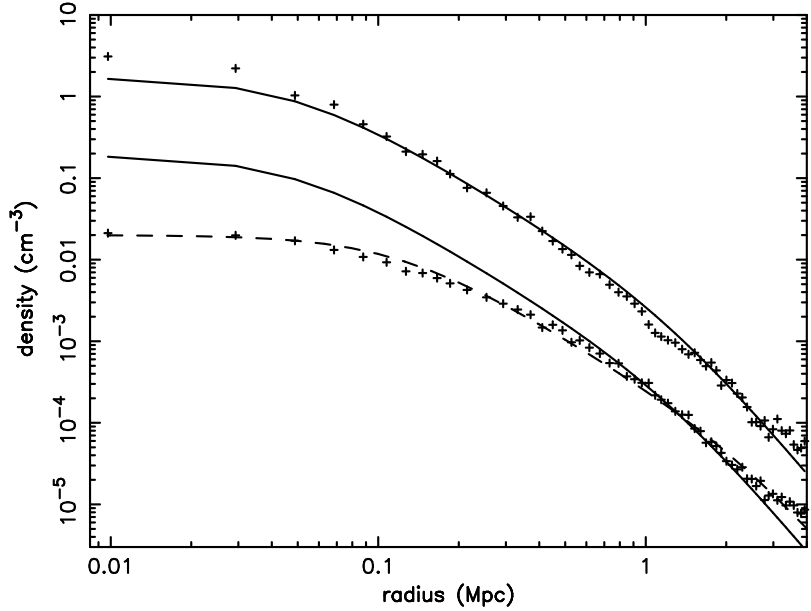

**Figure 11.** — The density profile for the dark matter and gas for the central cluster in the cluster simulation. The upper crosses are the dark matter and the lower for the gas. The solid lines are the fits derived in TC92 for the dark matter and gas profiles, of the form given in Equation 5. The lower dashed line is a fit of Equation 5 to the present gas density profile. The effective SPH smoothing length (the radius enclosing 32 particles) at the centre of the cluster is 0.1 Mpc.

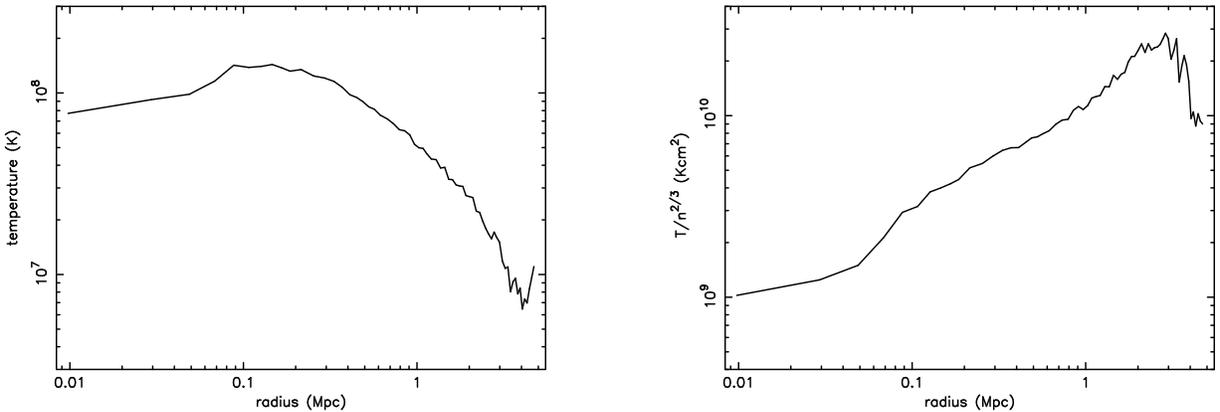

**Figure 12.** — (a) Temperature and (b) entropy profiles for the central cluster in the cluster simulation.

Why do the two codes give such different results? Both have been extensively tested and both perform well in shock-tube experiments to test the SPH. We attribute the differences principally to the different form of the energy equation, as discussed in Section 2.2. Both conserve energy but the older P³M+SPH form introduces erroneous entropy scatter. Particles in the low entropy tail of the distribution sink into the core of the cluster and give an artificially low temperature. If we replace the energy equation in the Hydra code with the old version, but keep everything else the same, then the core radius of the gas drops to $a = 80\,\text{kpc}$. Doubling the timestep results in a further decrease to about 70 kpc—we had to be more optimistic with the timestep in the P³M+SPH runs because of their much longer cpu time requirements. Small changes in other parameters, such as the shape of the SPH kernel or of the gravitational softening, lead to small fluctuations in the core radius about this value. This difference highlights the care that needs to be taken in order to avoid the introduction of unwanted entropy scatter in SPH simulations.

Our new simulations thus suggest that the conclusion of TC92, that the gas and dark matter have very similar dynamics, was premature. In particular the core radius of the gas distribution is much larger with the new code. This is consistent with the results of a series of merger simulations (Pearce, Thomas & Couchman 1994) in which we found that the dark matter could transfer energy to the gas, which then became irreversibly dissipated in shocks. Because the two-body relaxation time is so short in the core of these simulations, it is not possible to state with assurance that there should or should not be a core in the gas distribution. We intend to address this question in detail in a future paper based on much more detailed simulations.

## 8. SUMMARY AND CONCLUSIONS

In this paper we have described the algorithm and testing of Hydra — a hydrodynamics code which combines Adaptive $P^3M$ and Smoothed Particle Hydrodynamics. The adaptive nature of this scheme overcomes the major disadvantage of previous $P^3M$-SPH codes — that of severe slowing down under highly clustered conditions. The current code can follow the motion of 65 536 particles (half each of dark matter and gas) for 1000 timesteps in about 1 day on a Sun Sparc10 processor. Given that the cycle time of the algorithm scales essentially proportionately to the number of particles, a simulation with $10^6$ particles running for 1000 timesteps would require roughly 16 days to complete.

We have considered a variety of time integration schemes. In general low-order schemes with a single evaluation per timestep prove more stable and at least as accurate as Runge-Kutta type schemes for the same number of force evaluations per time interval. Our preferred choice is a single step scheme which is equivalent to Leapfrog for velocity-independent forces. A timestep constraint of $dt < \min(0.25dt_a, 0.5dt_c)$, where $dt_a$ and $dt_c$ are the acceleration and Courant timescales respectively, is sufficient to give excellent energy and entropy conservation.

A number of tests of the code were performed. We first showed that interparticle forces are correctly calculated with and without grid refinements. The code has been set up so that the *maximum* pairwise interparticle force may be accurately determined through a user selectable input parameter. A value of 7.7% was chosen which corresponds to an RMS value of $\simeq 1\%$. The RMS error in the net force on a particle in a typical distribution such as that found in a collapse simulation is of order 0.3%.

The SPH calculation was checked using a series of sound waves and shock tubes performed in a long cuboidal box with periodic boundary conditions. In each case a 'relaxed' initial particle distribution was used. This distribution was obtained by allowing SPH particles to spread out evenly with damped motion under the action of pressure forces alone. The residual power in such a distribution is very much less than that in white-noise and less than that in particle in cell initial conditions. Fluctuations in the measured SPH density of as little $\pm 1.3\%$ can be achieved. Such initial conditions are especially important for the tests described here for the following reasons: i) A distribution of particles on a cubical grid is unstable in SPH and ii) it is important in the SPH tests that particles are not aligned along the direction of propagation of sound waves or shocks as this restricts interpenetration and hence may lead to misleadingly optimistic results. Relaxed initial conditions of this kind are also a useful starting point for gravitational simulations for those who prefer to avoid grid starts as they have sufficiently low power up to the particle Nyquist frequency that many power spectra can be successfully imposed on the particle distribution without being unduly degraded by shot noise on small scales.

In both the sound wave and shock tests we can achieve excellent energy and entropy conservation with suitable choice of the timestep. Entropy conservation refers to maintaining a small scatter in the entropy of particles. This is a good test of SPH and is important in many simulations because, for example, entropy scatter provides a source of low entropy particles which may accumulate in potential wells and hence artificially raise the density there. Sound waves oscillate with approximately the correct frequency and are dispersed and damped with the kinetic energy being converted into heat by the artificial viscosity. The shock jump conditions are reproduced with reasonable accuracy with a shock-width of about 6 particles. The shortest timesteps give an entropy jump which is slightly too low due to incomplete thermalisation (longer timesteps do better). The RMS spread in the entropy is about 1% which we regard as good, but as this information is not available for other codes it is not possible to make a useful comparison with previous work. Energy is conserved to within 1% in all these runs.

We have also carried out investigations of collapsing spheres in an expanding uniform background. The tophat-collapse model for uniform spheres is reproduced extremely well provided that sufficient numbers of particles are used — the granularity of the particle distribution leads to a distortion of the surface layers which can propagate all the way into the centre. In practice many thousands of particles are required to model the collapse of such a smooth object with any degree of accuracy.

Energy conservation is tested for in the combined dark matter and gas distribution in the expanding simulation cube using a modified form of the Layzer–Irvine equation which allows for the thermal energy of the gas and for radiative cooling. This was applied to 'secondary infall' models in which the initial overdensity in a decreasing function of radius. Energy is conserved to within 0.3% during the collapse of a sphere containing about 10 000 particles, with the preferred integration scheme and timestep discussed above. Models containing dark matter alone do not collapse self-similarly, almost certainly because there is not sufficient dynamic range in the simulation. The gaseous collapse does seem to exhibit the correct self-similar behaviour in the core but once again the extent of this region is severely limited by the dynamic range available.

A more realistic test of the code was made by re-running a cluster simulation first performed using our previous, non-adaptive $P^3$M-SPH code. The dissipationless dark matter component behaves much as before but the gas now has a much more extended distribution. We attribute this difference mainly to excessive entropy scatter caused by a poor choice of energy equation in our old version of SPH. This highlights the importance of testing for both energy and entropy conservation in SPH simulations.

Our final remarks consist of a number of general comments on Hydra's main areas of applicability and on its strengths and weaknesses and the areas in which improvements can be made to the code.

The first point to emphasize is that although the code is clearly grid based, the ability to accurately reproduce sub-mesh-scale forces with the addition of a short range component to the mesh force in the hybrid $P^3$M scheme, renders the code essentially gridless in terms of particle dynamics. There are, however, a number of features of the present code which prevent it from being fully Lagrangian with the status that, for example, the Tree code enjoys. A variable smoothing length for the SPH particles has already been implemented and there is no impediment to doing the same for the gravitational interactions. The only limitation in this case is that the smoothing could not become larger than allowed by the base mesh force. This, however, is not a serious limitation given that we are primarily interested in particle dynamics in high density regions. Provision of multiple timestepping is another area which offers an improvement in the numerical model quite apart from the gain in efficiency that it offers. A simple form of multiple timestepping is being developed at present for AP$^3$M-type schemes in which a separate timestep is used at each level of refinement.

Although we anticipate that the code will find its main application in cosmological simulations in which the periodic boundary conditions of the base mesh are an asset it is certainly not limited to this area. Indeed we have used the code for simulations of objects with vacuum boundary conditions and one of us (FRP) has had considerable experience using this code as well as the Tree code for the simulation of isolated objects. The handling of particles moving to large distances from the main collapse is certainly problematic with a grid code but if the context in which a collapse occurs is at all important then errant particles will not be meaningful in either code.

Apart from its natural application to cosmological investigations the main strengths of the present code are that it is very fast and has a modest memory requirement. The performance figures at the beginning of this section show that it is possible to run a simulation with half a million particles for a thousand timesteps in a week on a typical workstation. With a memory requirement $\sim 20N$ words such a run would need 40Mb (assuming 4 byte words). This sort of performance is ideally suited to medium-resolution parameter space explorations. Since the cycle time of the code is proportional to the particle number the main impediment to larger simulations is the memory requirement. (Note that this is true of any code — the particle positions alone occupy $6N$ words.) On a modest 200 Mflop parallel machine with 80 M words of storage, 2 million particles each of gas and dark matter would take 3 days for 1000 steps.

We anticipate making Hydra available to the community within a few months.

## ACKNOWLEDGMENTS

The authors thank NATO for providing a Collaborative Research Grant (CRG 920182) to facilitate their interaction. The production of this paper was much aided by use of the STARLINK minor node at Sussex and by facilities supported by NSERC in Canada.